\documentclass[10pt, conference, compsocconf]{IEEEtran}
\usepackage{graphicx}
\usepackage{amsmath}
\usepackage{amsfonts}
\usepackage{amssymb}
\usepackage{subcaption}
\usepackage[ruled,linesnumbered]{algorithm2e}
\usepackage{algpseudocode}
\usepackage{multirow}
\usepackage[flushleft]{threeparttable}
\usepackage{tikz}
\usetikzlibrary{calc}
\usetikzlibrary{positioning}
\usetikzlibrary{arrows}
\tikzset{
    vertex/.style = {
        circle,
        fill            = black,
        outer sep = 2pt,
        inner sep = 1pt,
    },
    boxnode/.style = {align=center,draw,text width=2cm}
}
\tikzstyle{arrow}=[draw, -latex,solid,line width=0.2mm] 
\usepackage[noadjust]{cite}
\graphicspath{ {figs/} }
\usepackage{multicol}
\makeatletter
\def\BState{\State\hskip-\ALG@thistlm}
\makeatother

\begin{document}
%
% paper title
% can use linebreaks \\ within to get better formatting as desired
\title{Reliability Assessment and Quantitative Evaluation of Soft-Error Resilient 3D Network-on-Chip Systems}
\author{ \IEEEauthorblockN{Khanh N. Dang, Michael Meyer, Yuichi Okuyama, and Abderazek Ben Abdallah}
    \IEEEauthorblockA{The University of Aizu \\
        Graduate School of Computer Science and Engineering \\
        Aizu-Wakamatsu 965-8580, Japan \\
        Email: \{d8162103, d8161104, okuyama, benab\}@u-aizu.ac.jp}
}

% make the title area
\maketitle

\begin{abstract}
Three-Dimensional Networks-on-Chips (3D-NoCs) have been proposed as an auspicious solution, merging the high parallelism of the Network-on-Chip (NoC) paradigm with the high-performance and low-power cost of 3D-ICs. However, as technology scales down, the reliability issues are becoming more crucial, especially for complex 3D-NoC which provides the communication requirements of multi and many-core systems-on-chip. Reliability assessment is prominent for early stages of the manufacturing process to prevent costly redesigns of a target
system. 
In this paper, we present an accurate reliability assessment and quantitative evaluation of a soft-error resilient 3D-NoC based on a soft-error resilient mechanism. The system can recover from transient errors occurring in different pipeline stages of the router. Based on this analysis, the effects of failures in the network's principal components are determined.

\end{abstract}


% For peer review papers, you can put extra information on the cover
% page as needed:
% \ifCLASSOPTIONpeerreview
% \begin{center} \bfseries EDICS Category: 3-BBND \end{center}
% \fi
%
% For peerreview papers, this IEEEtran command inserts a page break and
% creates the second title. It will be ignored for other modes.
\IEEEpeerreviewmaketitle

\section{Introduction}
Global interconnects are becoming the largest performance bottleneck for high-performance Multi/Many-core Systems-on-Chip (MSoCs). For more than a decade, Network-on-Chip (NoC) interconnects have been proposed as a promising solution for future MSoC designs~\cite{Ben2006BNI}. The NoC paradigm offers more scalability than conventional shared-bus interconnects and allows more processing elements (PEs) to be efficiently integrated into a single chip. 
Despite the higher scalability and parallelism offered by a NoC system over traditional shared-bus based systems, it is still not an ideal solution for future large scale MSoCs. This is due to some limitations such as high power consumption and low throughput from NoCs with large dimensions. Taking NoCs to the third dimension %(3D-NoCs) 
has been proposed to deal with the above problems, as it was a solution offering lower power consumption and higher speeds~\cite{ Akram2013Aad, Akram2014gdf}. 

As feature sizes and supply voltages continually decrease, systems that have implemented these interconnects have become more vulnerable to soft errors. \textit{Shivakumar et al}.~\cite{sivakumar2002modeling} analyzed the transient error trends for smaller transistors and showed that the occurrence rate of transient faults is significantly higher than permanent faults. In particular, they expect the transient error rate for combinational logic to increase dramatically. Currently, soft error handling is mostly focused on memory and latches error which can be tackled by Error Correction Code. Therefore, future integration systems need a suitable solution to deal with soft errors. %In order to deal with soft errors, numerous works have proposed solutions in recent years. Soft errors are typically classified into two types of failures: failures in the data path and failures in the pipeline stages~\cite{Dang2015Softa}. 

Recently, numerous techniques have been proposed to handle soft errors~\cite{Dang2015Softa, Dang2016Soft, Yu2013Addressing, Prodromou2012NoCAlert} in NoCs; however, the reliability of the proposed techniques is still not well investigated. The most conventional method to evaluate the error resilience is injecting faults into the system and checking the output's accuracy. This kind of evaluation, especially at the gate level, requires a massive amount of time and computing resources. Moreover, the correction module is not sufficiently investigated in the reliability evaluation stage. Although the correction module helps to handle fault occurrences, it also suffers from an increase in fault probability. %This issue can lead to incorrect reliability results. 

In this paper, we present an efficient soft-error resilient mechanism and architecture for reliable 3D-NoC systems. We also present an accurate formulation of the reliability and vulnerability of the proposed 3D-NoC architecture against soft-errors. Our assessment method is based on the $Markov~state~model$~\cite{Shooman2003Reliability}. The fault-tolerance architecture is modeled as several states that include all of the possible failure cases. To switch between states, the $repairability$, and the failure rate is required. Based on the state model, a reliability function can be generated using probability functions. %Furthermore, we use the $Laplace~transform$ to help us calculate the MTTF (Mean Time to Failure) or MTBF (Mean Time Between Failure) of the system. 
Moreover, this paper presents a metric named the Reliability Acceleration Factor (RAF). This parameter is used to separate the reliability from the technology parameters, or operating conditions. By using the characteristics of the architecture and its fault-tolerance mechanism, the RAF value only represents the efficiency of the fault-tolerance mechanism alone. 

%%%%%%%%%%%%%%%%%%%%%%%%%%%%%%%%%%%%%%%%%%%%%%%%%%%%%
\section{System Architecture Overview} \label{sec:old-work}
%%%%%%%%%%%%%%%%%%%%%%%%%%%%%%%%%%%%%%%%%%%%%%%%%%%%%

\begin{figure}[hbtp]
    \centering
    \includegraphics[width=1\linewidth]{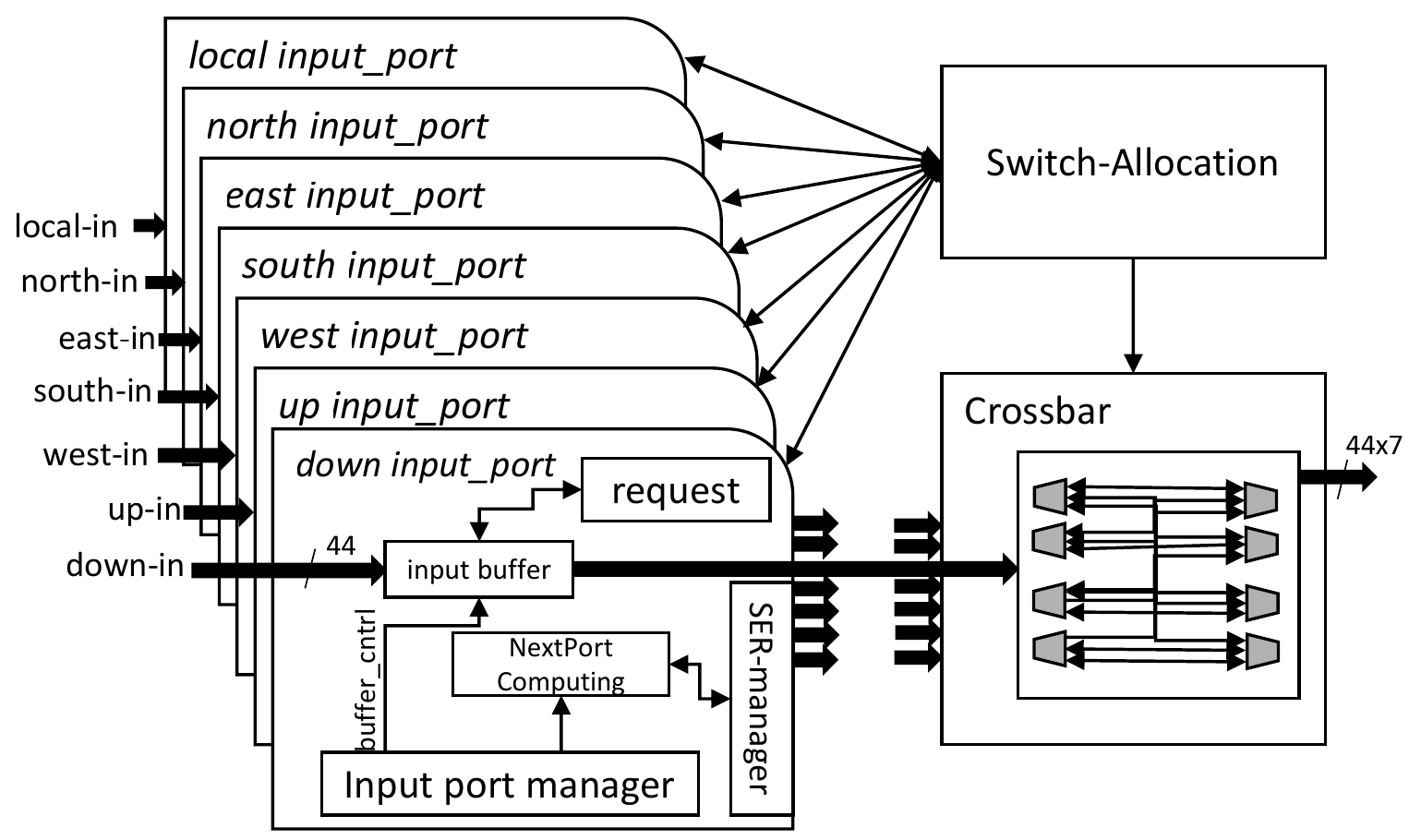}
    \caption{Proposed Soft Error Resilient 3D-NoC router architecture.}
    \label{fig:OASIS-arch}
\end{figure}

The 3D-Network-on-Chip (3D-ONoC) router block diagram is shown in Fig.~\ref{fig:OASIS-arch}. The router has three pipeline stages: (1) BW (Buffer Writing), (2) NPC/SA (Next Port Computation and Switch Allocation), and (3) CT (Crossbar Traversal). 
%The 3D-ONoC adopts a \textit{Wormhole-like} switching mechanism. 

The router architecture contains seven \textit{Input-port} modules for each direction in addition to the \textit{Switch-Allocation} and the \textit{Crossbar} module, which handle the transfer of flits to the next node. The\textit{ Input-port} module is composed of two main elements: \textit{Input-buffer }and the \textit{Next-Port-Routing} module. Incoming flits from the previous node or the local core, are first stored in the \textit{Input-buffer}. Since 3D-ONoC uses a \emph{Look-Ahead} routing algorithm, the output port value is already calculated by the previous node/core. Therefore, the \emph{output-port} value is sent with request signal to the \emph{Switch Allocation} module. At the same time, the \emph{Next-Port-Computing} module uses the routing information (\emph{destination} and \emph{output-port}) to calculate the routing i	nformation for the next node, labeled as \emph{next-output-port}. With the grant from \emph{Switch Allocation}, the flit will be sent through the \emph{Crossbar} to the next node together with the \emph{next-output-port} value.

%%%%%%%%%%%%%%%%%%%%%%%%%%%%%%%%%%%%%%%%%%%%%%%%%%%%%%%%%%%%%%%%%%%%%%%%%%%%%%%%%%%%%%%%%%%
\subsection{Soft-error Resilient Mechanism }\label{sec:SER-3DR}
%%%%%%%%%%%%%%%%%%%%%%%%%%%%%%%%%%%%%%%%%%%%%%%%%%%%%%%%%%%%%%%%%%%%%%%%%%%%%%%%%%%%%%%%%%%%5

Our main goal in proposing the SER-3DR (Soft-Error Resilient 3D-Network-on-Chip Router) is to develop a highly-reliable and low-cost technique 
to recover from soft-errors in all pipeline stages of the router.

For the errors in the pipeline stages, we propose a Soft Error Resilient Algorithm (SERA) , as shown in Algorithm~\ref{algo:SERA}. In the baseline OASIS-NoC, the router has three pipeline stages: BW (Buffer Writing), NPC/SA (Next Port Computing and Switch Allocation in parallel) and CT (Crossbar Traversal).
Since the NPC/SA stage (\textit{Routing and Arbitrating}) is where a majority of the complex combinational logic exists and it executes routing for the router, this stage is the one we selected for the SERA algorithm.

\begin{algorithm}[bhtp]
    \scriptsize
    \caption{Algorithm for SER-3DR}\label{algo:SERA}
    \tcp{input flit's data}
    \KwIn{in\_flit}
    
    \tcp{output flit's data}
    \KwOut{out\_flit}
    
    \vspace*{.2cm}
    
    \tcp{Write flit's data into buffers}
    
    \textbf{BufferWriting}(in\_flit)
    
    \tcp{Compute first time of NPC and SA}
    
    next\_port[1] = \textbf{NextPortComputing}(in\_flit)
    
    grants[1] = \textbf{SwitchAllocation}(in\_flit)
    
    \vspace*{.2cm}
    \tcp{ Compute redundant of NPC and SA}
    
    next\_port[2] = \textbf{NextPortComputing}(in\_flit)
    
    grants[2] = \textbf{SwitchAllocation}(in\_flit)
    
    \vspace*{.2cm}
    \tcp{Compare orginal and redundant to detect soft-error}
    
    \tcp{ Soft-error on NPC}
    
    \uIf {(next\_port[1] $\neq$ next\_port[2])} {
        \tcp{ roll-back and recalculate NPC}
        
        next\_port[3] = \textbf{NextPortComputing}(in\_flit)
        
        final\_next\_port =  \textbf{MajorityVoting}(next\_port[1,2,3]);
    }\Else{
    \tcp{ No soft-error on NPC}
    final\_next\_port = next\_port[1]    
}

\tcp{Soft-error on SA}
\uIf {(grants[1] $\neq$ grants[2])}{
    \tcp{roll-back and recalculate SA}
    grants[3] = \textbf{SwitchAllocation}(in\_flit)
    
    final\_grants = \textbf{MajorityVoting}(grants[1,2,3])
}\Else{ 
\tcp{No soft-error on SA}

final\_grants = grants[1]
}
\tcp{After detection and recovery, the algorithm finishes with CT}

out\_flit = \textbf{CrossbarTraversal}(in\_flit, final\_next\_port, final\_grants);

\end{algorithm}

As illustrated in Algorithm~\ref{algo:SERA}, SERA routes a flit from an input port to an output port. The input flit's data (in\_flit) is first written into the input buffer during the BW (Buffer Writing) stage (line 1). Second, SERA executes the first-time NPC (NextPortComputing) and SA (SwitchAllocation) stages which output the next\_port[1] and grants[1] respectively (lines 3-4). The NextPortComputing computes the routing path for the next node, similar to the look-ahead routing algorithm, and the SwitchAllocation handles the input port to output port routing. Third, the redundant processes of the NPC and SA are performed with these outputs: next\_port[2] and grants[2] (lines 4-5). In the next step, SERA compares the outputs of the original and the redundant processes. If next\_port[1] is different from next\_port[2], a soft-error occurred in the NPC, the algorithm then calculates the NPC a third time and uses majority voting to decide the final value (line 7-8). Otherwise, the final value is assigned as the first result (line 10). The SA is also processed in a similar fashion to the NPC: it starts by determining the occurrence of any errors, then voting on a value or assigning the first value (line 12-17). After detection and recovery, SERA finishes with crossbar traversal (line 18). The flit will be forwarded to the next node in the routing path or to the local core.

\section{Reliability Assessment}\label{sec:analysis}
\subsection{Mean Time Between Failure}
\begin{figure}[h]
    \centering
    \includegraphics[width=.7\linewidth]{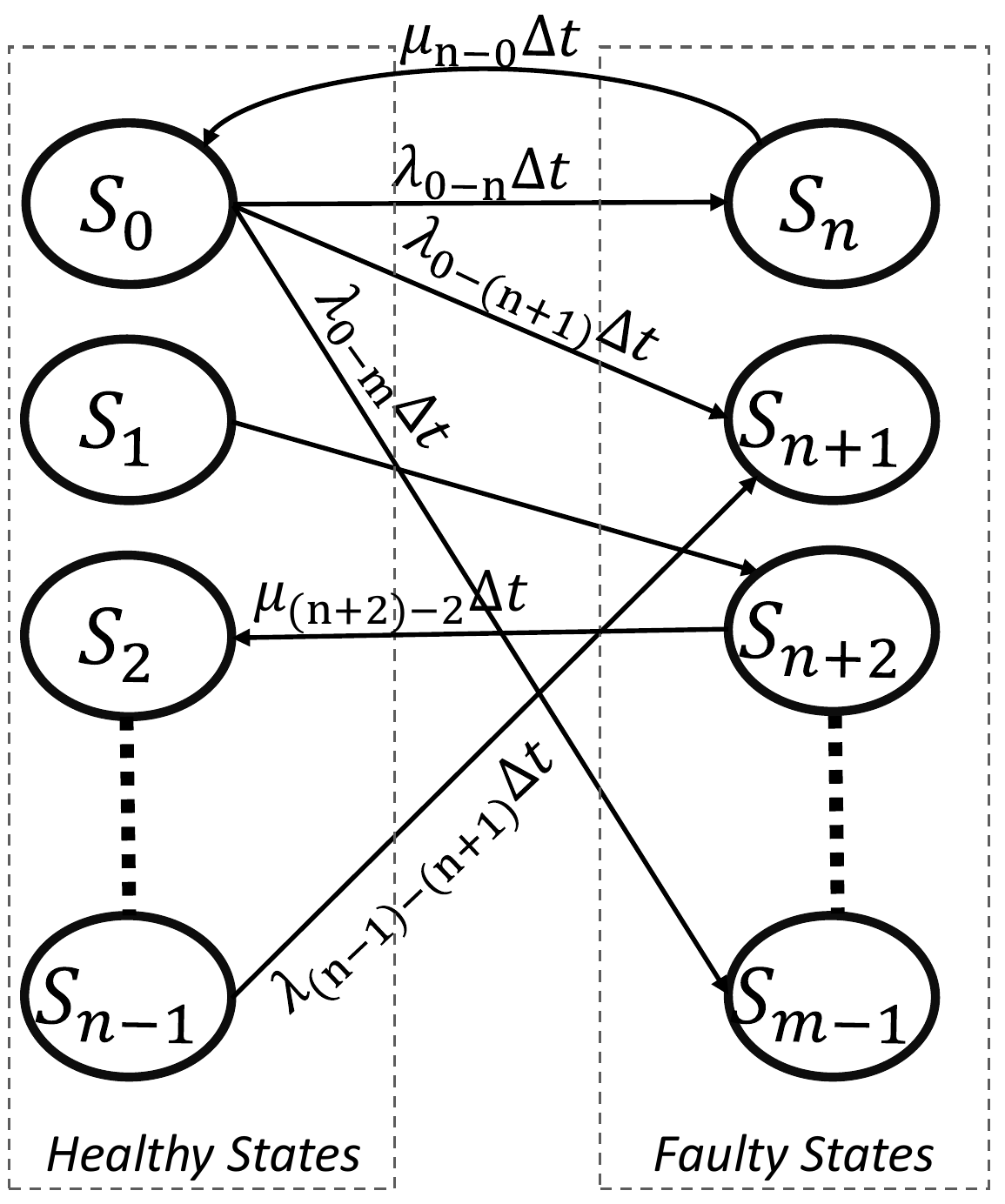}
    \caption{A Markov-state reliability model for a $m$ states system with $n$ non-failure states. } %, clock cycle: $5\ \mu s$, packet's size : 1-100 flits.}
    \label{fig:markov-model}
\end{figure}

In this section, we present a methodology to calculate the reliability of the system using the Markov State Model and the reliability function~\cite{Shooman2003Reliability}. The reliability of a system with respect to soft errors can be evaluated using the Mean-Time-Between-Failure calculation that follows:

\begin{equation} \label{eq:mtbf}
\mathbf{MTBF} = \lim_{s\to 0} R(s)
\end{equation}

where $R(s)$ is the reliability function of the system in the Laplace domain~\cite{Shooman2003Reliability}, which can be calculated based on the Markov state models as shown in Fig.~\ref{fig:markov-model}. Each state $S_i$ of the Markov model represents a possible status of the system. The status is defined as each case where an element of the system fails, which can lead to unpredictable operations. If the operation in $S_i$ state is correct, we define this state as ``healthy''. If the system is unable to operate correctly in state $S_i$, we define this state as ``faulty''.

Assuming the system has a set $\mathbb{S}$ which includes states from $S_0$ to $S_m$ in its Markov state model. We use $S_0$ to represent the initial state of the system. The set of healthy states and set of faulty states are defined as:

\begin{equation}
\mathbb{H} \triangleq \{S_i \in \mathbb{S} |\mathrm{ the\ system\ works\ correctly\}}
\end{equation}

\begin{equation}
\mathbb{F} \triangleq \{S_i \in \mathbb{S} |\mathrm{  the\ system\ not\ working\}}
\end{equation}

where:
\begin{itemize}
    \item $\mathbb{H}$ is the set of states which still maintain the operation of the system, %We assume this set has $n$ elements.
    \item $\mathbb{F}$ is the set of states which the system fails. 
\end{itemize}
The reliability function $R(s)$ in Equation~\ref{eq:mtbf} is defined as follows:

\begin{equation}
R(s) = P(\mathbb{H}) = 1 -  P(\mathbb{F})
\end{equation}

As shown in Fig.~\ref{fig:markov-model}, each state $S_i$ represents a status of the system during operation. There are two subsets,  healthy states $H = \{S_0, S_1, ...,S_{n-1}\}$ and faulty states $F = \{S_n, S_{n+1}, ...,S_{m-1}\}$. The transition rate are defined as follows:
\begin{itemize}
    \item $\lambda$ is the fault rate a component in the system which can represent for a transition from a element of set $\mathbb{H}$ to a element of set $\mathbb{F}$. 
    \item $\mu$ is the repair rate a component in the system which can represent for a transition from a element of set $\mathbb{F}$ to the element of set $\mathbb{H}$.  
\end{itemize}

The final MTBF can be calculated based on the healthy states as:

\begin{equation}  \label{eq:MTBF-1}
\mathbf{MTBF} = \lim_{s\to 0} (P(\mathbb{H}(s)))   = \lim_{s\to 0} \sum_{i=0}^{n-1}{S_i(s)}
\end{equation}

%or based on the faulty states: 
%
%\begin{equation}  \label{eq:MTBF-2}
%\mathbf{MTBF} = \lim_{s\to 0} (1-P(\mathbb{F}(s)))   = 1 - \lim_{s\to 0} \sum_{i=n}^{m-1}{S_i(s)}
%\end{equation}

\subsection{Fault Model}  \label{ssec:fault-assumption}
Before analyzing the reliability of the system and calculating the MTBF, we need make two assumptions about the failure of the system.
\begin{itemize}
    \item {The system starts with a default all components healthy state}. In Fig.~\ref{fig:markov-model}, the initial status is: $S_0 = 1$ and $S_i = 0$ with $i\neq 0$.
    \item {The transition rate $\lambda$ and $\mu$ are constants and per-determined}. 
\end{itemize}

However, the fault rate and repair rate depend on the technology parameters, running environment and operating circumstances. In order to separate them from the fault rate value, we propose these assumptions:
\begin{itemize}
    \item {The fault rate has a linear relationship to the area cost of the module.} 
    \item {The fault rate has a linear relationship to the operating time of the module.} 
    \item {The fault rate is affected after the module is attached to a system.} 
\end{itemize}           
In this fashion, the fault rate only depends on the area cost, the operating time and the efficiency of the reduction of the fault rate that is provided by the fault-tolerance mechanism. Thus, for a system with $k$ components, its fault rate is given by:

\begin{equation}   \label{eq:change-fit}
\lambda_{system} = \sum_{i=1}^{k}f_i \mathbf{OR}_{i}\mathbf{AR}_{i}\lambda_{unit}
\end{equation}     

where: 
\begin{itemize}
    \item $unit$ is a select module as a reference for calculation. In our method, we use the original design as a $unit$.
    \item $\mathbf{OR}_i$ is the operating time ratio of component $i$ to $unit$.
    \item $\mathbf{AR}_i$ is the area cost ratio of component $i$ to $unit$.
    \item $f_i$ is the changing rate caused by attaching the module $i$ to the system. If the system has no impact to the failure of module $i$ or module $i$ is standalone, $f_i = 1$. If the system has a fault-tolerance mechanism for module $i$, $f_i < 1$. 
\end{itemize}

Soft errors occur over a short period of time, typically within a single clock cycle. Because of this, the combinational logic has the ability to self-correct soft errors. On the other hand, the memory units (latch, flip-flop,...) still record failures. %Therefore, we define the repair rate for soft errors as:

\subsection{Reliability Evaluation}

In order to evaluate a fault-tolerance method, we introduce a parameter named the Reliability Acceleration Factor (RAF), which is defined as:
\begin{equation}  \label{eq:raf}
\mathsf{RAF} = \frac{\mathbf{MTBF}_{FT}}{\mathbf{MTBF}_{original}} 
\end{equation}  
where ${\mathbf{MTBF}_{FT}}$ and ${\mathbf{MTBF}_{original}}$ are \textit{Mean Time Between Failures} of the Fault-Tolerant and the original system, respectively.

Based on analysis of the operation of the original and fault-tolerant system, we can obtain both MTBFs using Equation~\ref{eq:MTBF-1}. Moreover, by using Equation~\ref{eq:raf}, the RAF parameter is separate from the fault rate of the $unit$ module. In this fashion, the RAF parameter only represents the efficiency of the fault-tolerance mechanism. 

\begin{figure}[h]
    \centering
    \includegraphics[width=.9\linewidth]{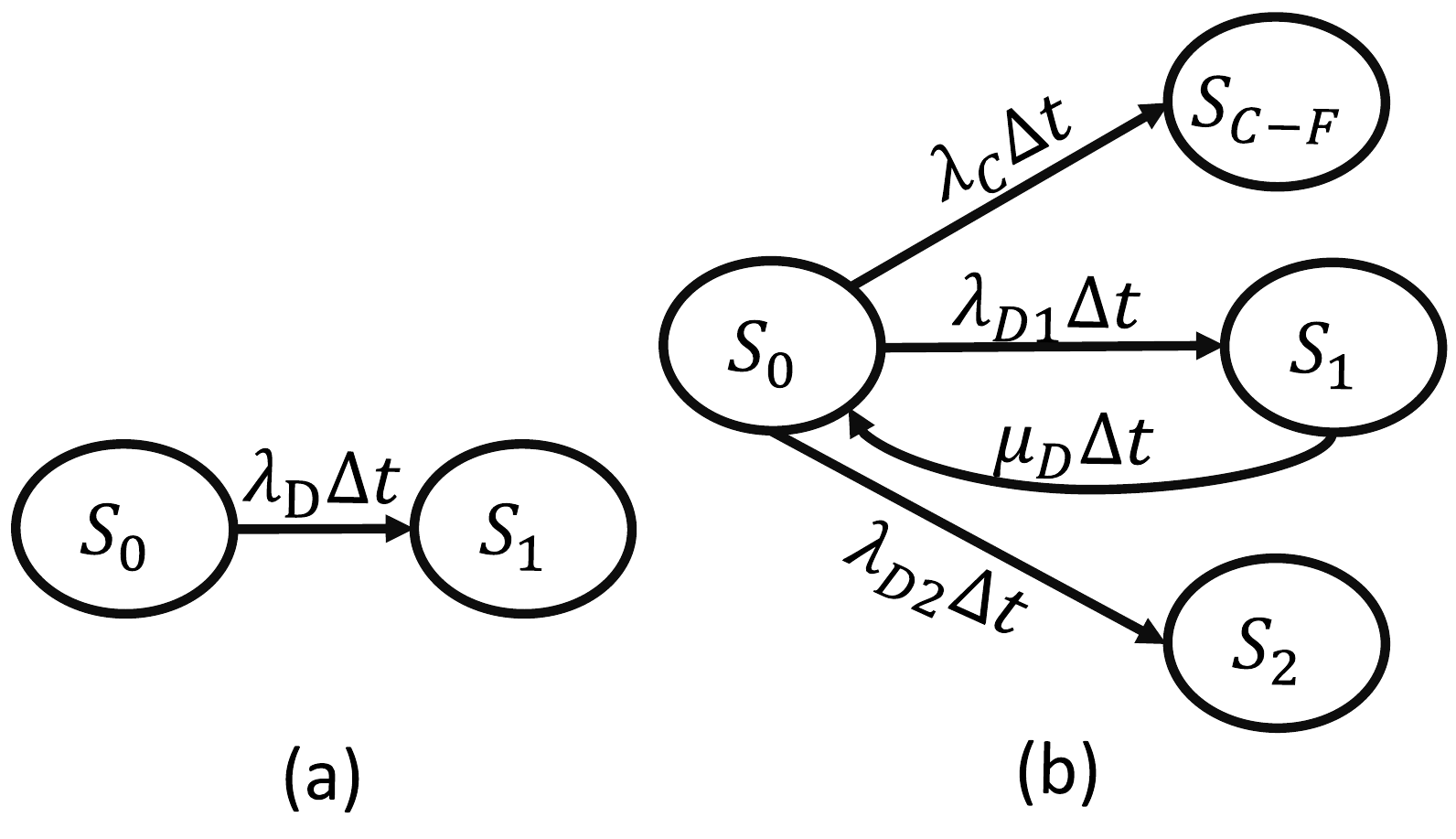}
    \caption{A simplified Markov-state reliability model for: (a) the original system; (b) the fault-tolerant (FT) system.} 
    \label{fig:markov-protected}
\end{figure}  

In order to evaluate a fault-tolerant system which consists of both original modules and correction modules, we model the fault-tolerant system to have a Markov state similar to Fig.~\ref{fig:markov-protected} (b) with:
\begin{itemize}
    \item $S_0$ is the initial state.
    \item $S_1$ in Fig.~\ref{fig:markov-protected} (b) is the failure state of the original system which is given by a transition rate $\lambda_{D1}$.
    \item $S_2$ in Fig.~\ref{fig:markov-protected} (b) is the failure state of the original system which is given by a transition rate $\lambda_{D2}$.
    \item $S_{C-F}$ in Fig.~\ref{fig:markov-protected} (b) is the failure state of the repair module.% system which is given by a transition rate $\lambda_D2$.
\end{itemize}
Because of the protection from the fault-tolerance technique, the FT system can handle some kinds of faults. Therefore, we define the transition rates as:
\begin{itemize}
    \item $\lambda_D$ is the fault rate of the original system (D).
    \item $\lambda_C$ is the fault rate of the repair module of the FT system.
    \item $\mu_D$ is the repair rate which is provided by the repair module (C).
    \item $\lambda_{D1}$ is the part of the fault rate in the original which is handled by the repair module (C).
    \item $\lambda_{D2}$ is the part of the fault rate in the original which the repair module (C) cannot correct.
\end{itemize}
Based on the Markov state of the two systems, as shown in Fig.~\ref{fig:markov-protected}, the MTBFs are given as:  

\begin{equation}  \label{eq:ori}
\mathbf{MTBF}_{original} = \frac{1}{\lambda_D}
\end{equation}   
and 
\begin{equation}  \label{eq:ft}
\mathbf{MTBF}_{FT} =  \frac{1}{\lambda_{D2} + \lambda_C}
\end{equation}

By applying Equations~\ref{eq:change-fit},~\ref{eq:ori}, and~\ref{eq:ft} to Equation ~\ref{eq:raf}, we have:
\begin{equation}  \label{eq:final-eq}
%\scriptsize
\mathsf{RAF} = \frac{1}{(f_D\times \mathbf{OR}_D\times \mathbf{AR}_D) + (\mathbf{OR}_{C} \times \mathbf{AR}_{C})}
\end{equation} 
where:
\begin{itemize}
    \item $f_D = \lambda_{D2}/\lambda_{D}$ is the ratio of the failure rate of the original system after and before applying the fault-tolerance mechanism. 
    \item $\mathbf{OC}_{x}$  is the ratio of operation of the module $x$ to the original system (D).
    \item  $\mathbf{AR}_{x}$ is the ratio of area cost of the module $x$ to the original system (D).
\end{itemize}

Equation~\ref{eq:final-eq} shows the RAF function based on the architecture modifications in area cost and operation and the reduction of the failure rate.

%%%%%%%%%%%%%%%%%%%%%%%%%%%%%%%%%%%%%%%%%%%%%%%%%%%%
\section{Design and Evaluation Results}\label{sec:res}
%%%%%%%%%%%%%%%%%%%%%%%%%%%%%%%%%%%%%%%%%%%%%%%%%%%%

%
Our proposed system (SER-3DR) is integrated into the 3D-ONoC~\cite{Akram2014gdf,Akram2013Aad}. We designed the system in Verilog-HDL, and synthesized it using the 45nm technology library. %~\cite{nangate2014nangate}. For the Through-Silicon-Via (TSV) integration, we used the FreePDK3D45 kit compiler~\cite{NCSUEDA2015FreePDK3D45}.
We evaluated the hardware complexity, power consumption and speed. 
We also evaluated performance using three benchmarks:  Matrix-multiplication, Transpose, and Uniform were selected.
For comparison, we also implemented and simulated the baseline LAFT-OASIS ~\cite{Akram2013Aad}, and Triple Modular Redundancy of the NPC/SA based on OASIS (TMR)~\cite{Dang2015Softa}. % as mentioned Fig.~\ref{fig:MJV}. 
In order to demonstrate the reliability, we analyze our architecture to provide the RAF parameter in Eq.~(\ref{eq:final-eq}) and provide some comparison with another mechanism.

\subsection{Hardware Complexity} 
Table~\ref{tab:imple} depicts the implementation result of a single router of the original OASIS system, the TMR router~\cite{Dang2015Softa}, and the proposed SER-3DR router. The router is designed for a 3D Mesh with 7 ports, 32 bit flit-width, Stall-Go flow control and wormhole forward mechanism.
When compared with the original LAFT-OASIS router architecture, the SER-3DR requires slightly more logic area cost~($14.98\%$). The TMR router costs $45.20\%$ more because it triplicates the NPC and SA stage. The frequency decreases from $801.28\ MHz$ to $655.74\ MHz$ ($-18.16\%$).
Also, the TMR system increases the power consumption to $30.31\ mW$ ($+18.30\%$). The proposed design slightly increases the power consumption from the baseline's $25.62\ mW$  to $27.13\ mW$ ($+5.90\%$).  

\begin{table}[tbh]
    \scriptsize
    \begin{center}
        \caption{Hardware complexity comparison results.} \label{tab:imple}
        \begin{tabular}{|c|c|c|c|c|}
            \hline
            \multirow{2}{*}{\textbf{Design}} &\textbf{ Max Freq.}& \textbf{Total Power}  &  \textbf{Logic's area} & \textbf{\# TSVs} \\
            & \textbf{(MHz) }& \textbf{$mW$)} & \textbf{($\mu m^2$)} & \\  \hline \hline
            LAFT OASIS & 801.28  & 25.62  & 14,920 & 164 \\ \hline
            TMR-OASIS & 763.36 & 30.31 & 21,664 & 164 \\ \hline 
            SER-3DR & 655.74  & 27.13	 & 17,154 & 164\\ \hline
            
        \end{tabular}
    \end{center}
\end{table}    

%%%%%%%%%%%%%%%%%%%%%%%%%%%%%%%%%%%%%%%%%%%%%%5
\subsection{Network Performance Evaluation} \label{sec:eva-time}
%%%%%%%%%%%%%%%%%%%%%%%%%%%%%%%%%%%%%%%%%%%%%%%%
\begin{figure*}[bhtp]
    \begin{subfigure}{0.5\textwidth}
        \centering
        \includegraphics[width=.9\linewidth]{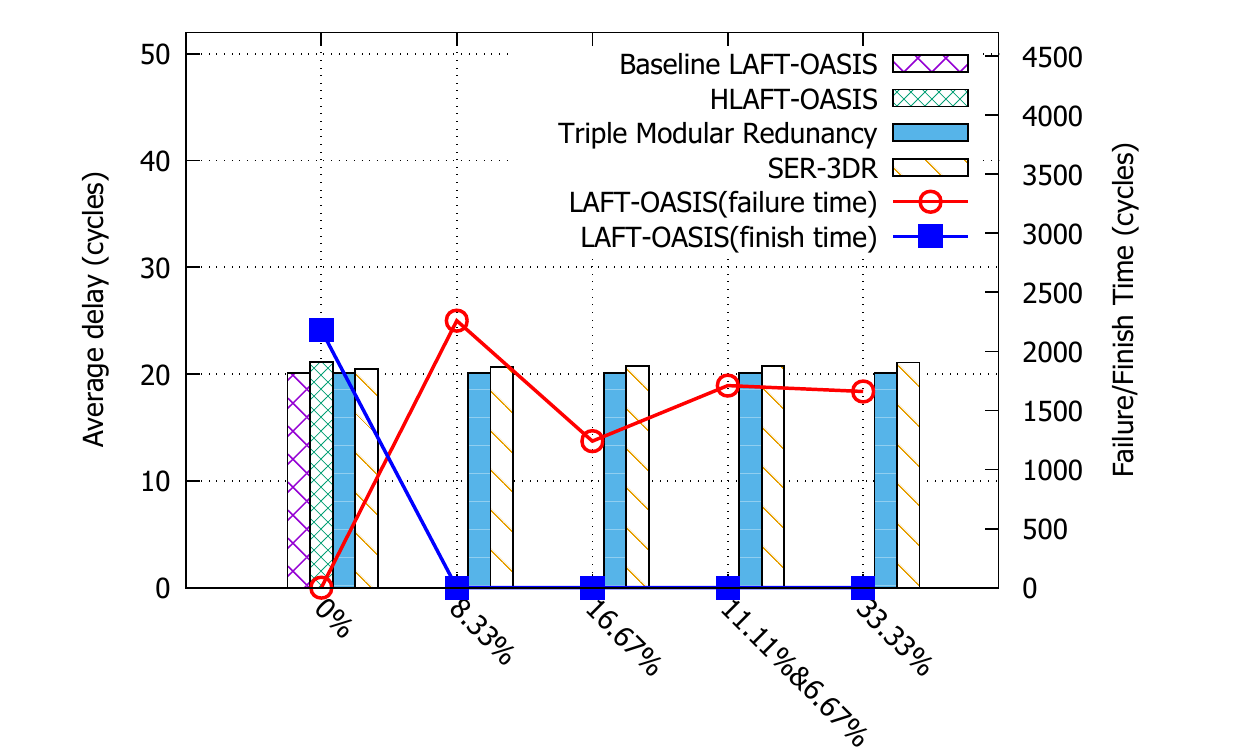}
        \caption{Average Delay of Transpose.}
    \end{subfigure}
    \begin{subfigure}{0.5\textwidth}
        \centering
        \includegraphics[width=.9\linewidth]{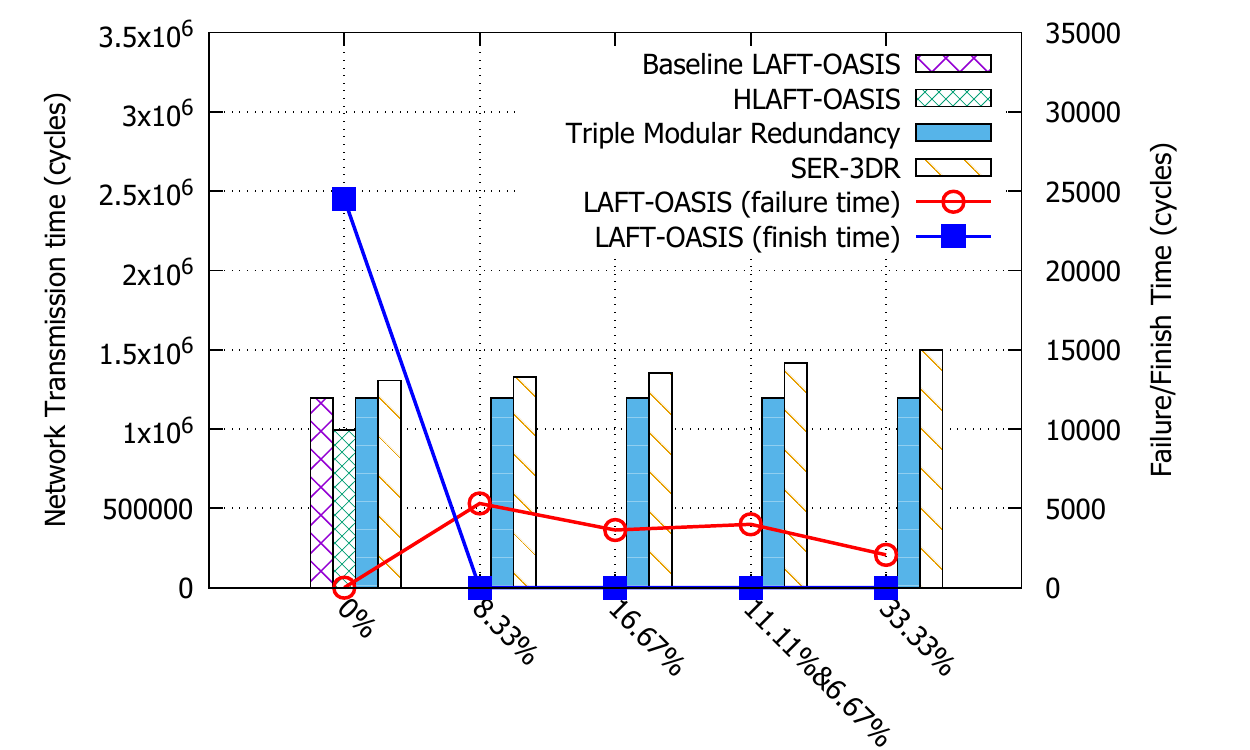}
        \caption{Network Transmission Time of Uniform. }
    \end{subfigure}
%    \begin{subfigure}{0.33\textwidth}
%        \centering
%        \includegraphics[width=\linewidth]{matrix_compare.pdf}
%        \caption{Network Transmission Time of Matrix.}
%    \end{subfigure}  
    \begin{subfigure}{0.5\textwidth}
        \centering
        \includegraphics[width=.85\linewidth]{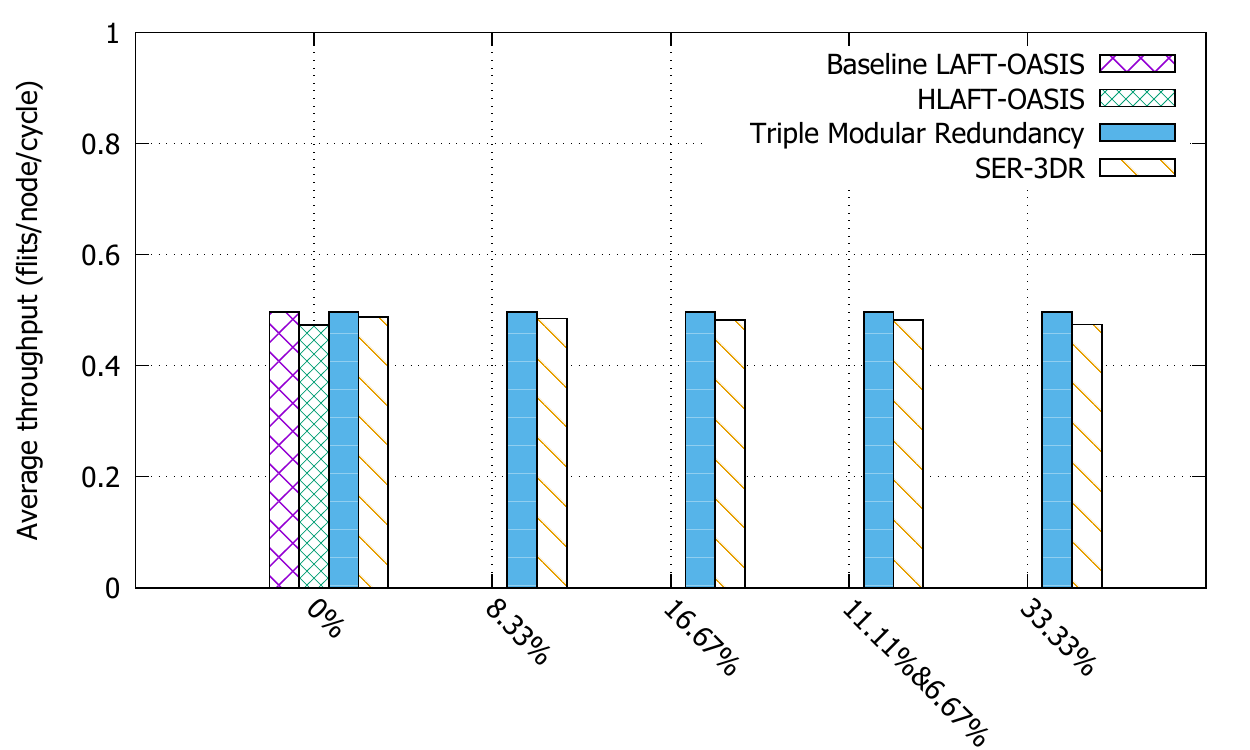}
        \caption{Throughput of Transpose.}
    \end{subfigure}
    \begin{subfigure}{0.5\textwidth}
        \centering
        \includegraphics[width=.85\linewidth]{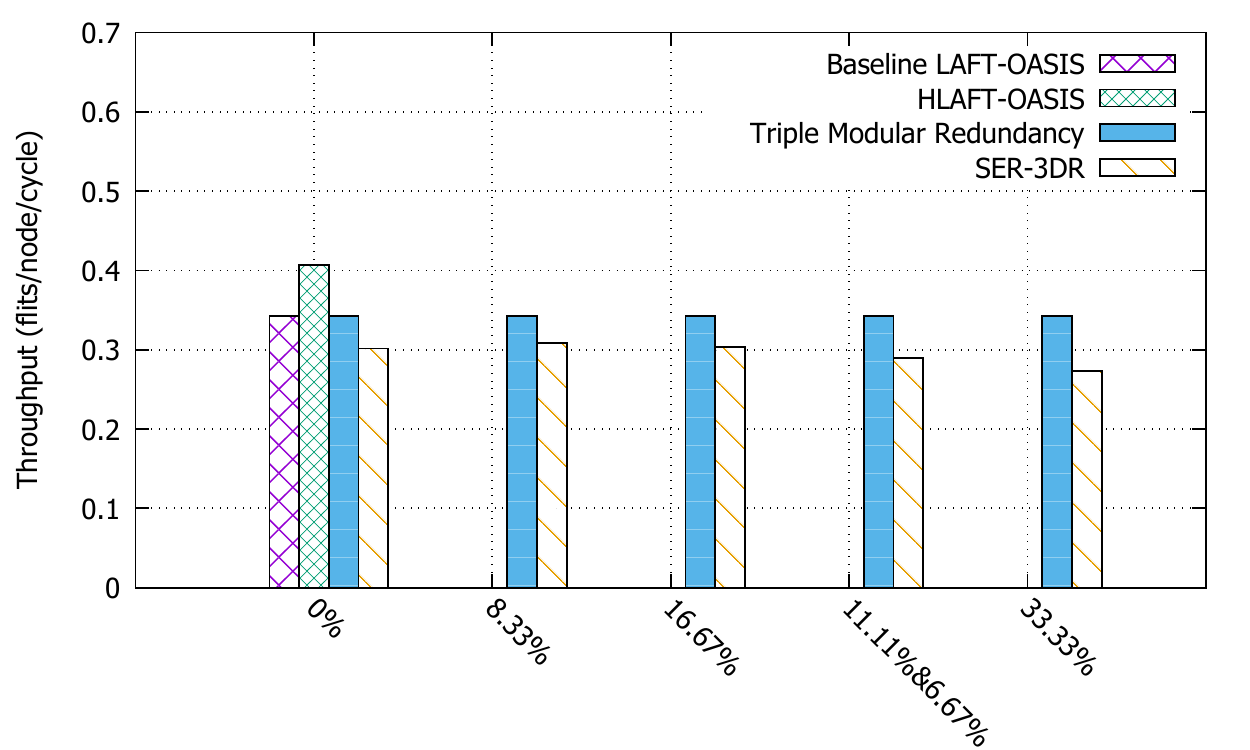}
        \caption{Throughput of Uniform.}
    \end{subfigure} 
%    \begin{subfigure}{0.33\textwidth}
%        \centering
%        \includegraphics[width=.9\linewidth]{matrix_throughput.pdf}
%        \caption{Throughput of Matrix.}
%    \end{subfigure} 
    \caption{Simulation's Results.}
    \label{fig:sim-results}
\end{figure*}

For this evaluation, we used the three benchmarks over five injection rates : $0\%$, $8.33\%$, $16.67\%$, $11.11\%\&6.67\%$ (in Routing and Switch Allocator) and $33\%$. 
%
%For Transpose, we measured the average delay time as execution time. 
The evaluation results with Transpose and Uniform are shown in Fig.~\ref{fig:sim-results} (a and b), respectively. 
We perform these benchmarks for 3 models (SER-3DR, LAFT-OASIS and the TMR). The network transmission time or average delay is presented as a bar graph. We also inject the soft-errors inside the baseline model (LAFT-OASIS) and measure the transmission time. Its \textit{failure time} and complete \textit{finish time} are depicted in line graph format.

For the Transpose benchmark in Fig.~\ref{fig:sim-results}(a), we found that the average latency slightly increased from $20.113\ cycles s$ to $20.505\ cycles$ ($+1.95\%$) for an error injection rate of  $0\%$. With different error injection rates, we can see that the average latency slightly increased from $20.505\ cycles$ for a $0\%$ error rate to  $21.092$ $cycles$ for a $33\%$ error rate. Uniform benchmark has a $9.06\%$ increase in transmission time with an absence of faults, while Matrix has a $10.02\%$ additional transmission time. In the faulty cases, SER-3DR requires additional intervals for detecting and recovery.

To perform the throughput evaluation, we also used the above benchmarks with five injection rates as shown in Fig.~\ref{fig:sim-results} (c and d). For the Uniform benchmark, the throughput is slightly degraded due to the short packet length which increases the impact of redundant cycles.

\subsection{Reliability Comparison}

\begin{table*}[thbp]
    \begin{center}
        \scriptsize
        \caption{Comparison of soft-error resilient mechanisms for routing and arbitrating modules in Network-on-Chip router.}
        \label{tb:soft-compare}
        \begin{tabular}{|p{2cm}||p{2.5cm}|p{2.5cm}|p{2.5cm}|p{2.5cm}|}
            \hline
            
            \textbf{Model}       & TMR for OASIS~\cite{Dang2015Softa}  & Yu et al.~\cite{Yu2013Addressing} & Prodromou et al.~\cite{Prodromou2012NoCAlert} & Our proposal~\cite{Dang2015Softa}  \\ \hline  \hline
            Mechanism & Majority Voting & Monitor & Monitor & Monitor \\ \hline
            Area Overhead & 204.33\% & 9\% & 3\% (average) & 54.46\% \\ \hline
            $\mathbf{AR}_{C}$  & 0.0433 & 0.09 & 0.03  & 0.5446 \\ \hline
            $\mathsf{RAF}$ &   $\simeq 1.33$ &  $\simeq 11.11$ &   $\simeq 1$ (only detection)   & $ 1.84$\\  \hline
            Delay       & +0 &  +0 cycle (no fault) & 0\% (only detection) & +1 cycle (redudancy) \\
            &     &  +1 cycle (recovery) &                   & +2 cycle (recovery) \\ \hline
            Fault Coverage & 100\% of permanent & design specific  & design specific  & 100\% transient \\
            &  and transient  & (7 faults) & (13 faults) &  \\ \hline
            %Design Modification & light & medium & only detector & medium (FSM state) \\ \hline
            Reovery method & immediately & re-execution & unsupport & re-execution \\ \hline
        \end{tabular}
    \end{center}
\end{table*} 

In this section, we compare the soft error resilience technique designed for the two modules: Routing and Switch Allocation. We select the TMR and two architectures from~\cite{Prodromou2012NoCAlert,Yu2013Addressing}. The architecture by Prodromou et al.~\cite{Prodromou2012NoCAlert} and Yu et al.~\cite{Yu2013Addressing} use monitors to verify the accuracy of the output values. To verify the output value, they predefine a set of rules based on the architecture of the monitored modules. If the output value violates one of the rules, the monitor module can determine the failure and decide the next step. The architecture by Yu et al.~\cite{Yu2013Addressing} allows the system to re-execute the task again to correct the failure. Prodromou et al.~\cite{Prodromou2012NoCAlert} does not provide any recovery method for the detected failure. Although the set of faults for the routing and arbitrating modules consists of 7 and 13 faults, we assume both architectures can handle 100\% of transient faults in this evaluation.

%According to~\cite{Shooman2003Reliability}, the reliability function of TMR is:
%\begin{equation}
%R(s) = \frac{s+5\lambda_D + \mu}{s^2+(5\lambda_D+\mu)s+ 6\lambda_D^2}
%\end{equation} 

For our proposal~\cite{Dang2015Softa} and Yu et al.~\cite{Yu2013Addressing}, we use Eq.~\ref{eq:final-eq} to calculate the RAF value. The reliability function of TMR is obtained from~\cite{Shooman2003Reliability}. The original module is not modified to implement the fault-tolerance architecture. Therefore, the area cost and operating ratios ($AR_D$ and $OR_D$) of the protected module both equal 1. Since the operation of the correction module follows the operation of the protected module, the operating ratios of the correction module $OR_C$ are both equal to 1. The area ratios $AR_C$ of the correction module are shown in Table~\ref{tb:soft-compare}. Since our proposal can handle 100\% kind of the error, we assume that Yu et al.~\cite{Yu2013Addressing} also has the same capacity.

As shown in Table~\ref{tb:soft-compare}, our proposed architecture has a medium area cost overhead (54.46\%). Both monitor based architectures have a small area overhead (9\% and 3\%). The TMR architecture has the largest area cost (+204.33\%) but it has the least impact on performance and architecture modification. Although monitor based architectures provide a small area overhead, they only cover the faults which violate their set of rules. Therefore, a new architecture requires another set of rules to handle soft errors. On the other hand, our proposal and TMR can cover 100\% kind of transient fault and adapt to any architecture without significant modifications. 

In terms of RAF (Reliability Acceleration Factor), the TMR attains a value of around 1.33. %According to Shooman et. al~\cite{Shooman2003Reliability}, TMR is superior within early time of operation. 
Our architecture provides a RAF of 1.84. For monitor-based architectures, Yu et al.~\cite{Yu2013Addressing} provides the best protection with 11.11; however, this value is attained under the assumption that the design will cover 100\% of failures. The architecture by Prodromou et al.~\cite{Prodromou2012NoCAlert} does not improve due to the fact that they did not implement a recovery module. 

% % % % % % % % % Conclusion % % % % % % % % % % %
\section{Conclusion}
In this paper, we presented an accurate reliability assessment and quantitative evaluation of a soft-error resilient 3D-NoC system based on a soft-error resilient mechanism. The system can recover from transient errors occurring in different pipeline stages of the router. Evaluation results show that the system has about 1.84 times improvement in MTBF while handling $100\%$ of all tested error types. Moreover, the system can achieve a high level of transient error protection with a small latency increase of $18.16\%$ when compared to the baseline non-protected system.  
%\\
%\\
%\textbf{Acknowledgment:}  
%This work is supported by VLSI Design and Education Center (VDEC), the University of Tokyo, Japan, in collaboration with Synopsis, Inc. and Cadence Design Systems, Inc. This project is partially supported by Competitive Research Funding of the University of Aizu, Fukushima, Japan, Ref. P11-2016. 


\begin{thebibliography}{1}

    \bibitem{Ben2006BNI}
    A.~B. Abdallah and M.~Sowa, ``{Basic Network-on-Chip Interconnection for Future
      Gigascale MCSoCs Applications: Communication and Computation
      Orthogonalization},'' in {\em Proc. of the Symposium on Science, Society, and
      Technology (TJASSST2006)}, pp.~1--7, 2006.
    
    \bibitem{Akram2013Aad}
    A.~B. Ahmed and A.~B. Abdallah, ``{Architecture and design of high-throughput,
      low-latency, and fault-tolerant routing algorithm for 3D-network-on-chip
      (3D-NoC)},'' {\em The Journal of Supercomputing}, vol.~66, no.~3,
      pp.~1507--1532, 2013.
    
    \bibitem{Akram2014gdf}
    A.~Ben~Ahmed and A.~Ben~Abdallah, ``{Graceful deadlock-free fault-tolerant
      routing algorithm for 3D Network-on-Chip architectures},'' {\em Journal of
      Parallel and Distributed Computing}, vol.~74, no.~4, pp.~2229--2240, 2014.
    
    \bibitem{sivakumar2002modeling}
    P.~Sivakumar, M.~Kistler, S.~Keckler, D.~Burger, and L.~Alvisi, ``Modeling the
      effect of technology trends on soft error rate of combinatorial logic,'' in
      {\em Proc. Intl. Conf. Dependable Sys. \& Networks DSN�02}, pp.~23--26, 2002.
    
    \bibitem{Dang2015Softa}
    K.~N. Dang, M.~Meyer, Y.~Okuyama, A.~Ben~Abdallah, and X.-T. Tran, ``Soft-error
      resilient 3d network-on-chip router,'' in {\em Awareness Science and
      Technology (iCAST), 2015 IEEE 7th International Conference on}, pp.~84--90,
      IEEE, Sep 2015.
    
    \bibitem{Dang2016Soft}
    K.~N. Dang, Y.~Okuyama, and A.~B. Abdallah1, ``Soft-error resilient
      network-on-chip for safety-critical applications,'' in {\em 2016
      International Conference on IC Design and Technology (ICICDT)}, pp.~1--4,
      June 2016.
    
    \bibitem{Yu2013Addressing}
    Q.~Yu, M.~Zhang, and P.~Ampadu, ``Addressing network-on-chip router transient
      errors with inherent information redundancy,'' {\em ACM Transactions on
      Embedded Computing Systems (TECS)}, vol.~12, no.~4, p.~105, 2013.
    
    \bibitem{Prodromou2012NoCAlert}
    A.~Prodromou, A.~Panteli, C.~Nicopoulos, and Y.~Sazeides, ``Nocalert: An
      on-line and real-time fault detection mechanism for network-on-chip
      architectures,'' in {\em Microarchitecture (MICRO), 2012 45th Annual IEEE/ACM
      International Symposium on}, pp.~60--71, Dec 2012.
    
    \bibitem{Shooman2003Reliability}
    M.~L. Shooman, {\em Reliability of computer systems and networks: fault
      tolerance, analysis, and design}.
    \newblock John Wiley \& Sons, 2003.
    
    \end{thebibliography}
\end{document}